\documentclass[numbered]{trbunofficial}
\pdfoutput=1
\usepackage{graphicx}
\usepackage{booktabs}
\usepackage{subcaption}
\usepackage{multirow}

\usepackage{hyperref}

\usepackage{booktabs}

\AuthorHeaders{Zhu, Hu, Yang, Pu, and Wang}
\title{Personalized Context-Aware Multi-Modal Transportation Recommendation}

\author{%
   \textbf{Meixin Zhu}\\
  Research Assistant\\
  Department of Civil and Environmental Engineering\\
 University of Washington\\
    Seattle, WA 98195-2700\\
  meixin92@uw.edu\\
  \hfill\break
  \textbf{Jingyun Hu}\\
  Master Student\\
  Department of Civil and Environmental Engineering\\
 University of Washington\\
    Seattle, WA 98195-2700\\
  jingyun@uw.edu\\
  \hfill\break
  \textbf{Hao (Frank) Yang}\\
  Research Assistant\\
  Department of Civil and Environmental Engineering\\
 University of Washington\\
    Seattle, WA 98195-2700\\
  haoya@uw.edu\\
  \hfill\break
  \textbf{Ziyuan Pu}\\
  Research Assistant\\
Department of Civil and Environmental Engineering\\
 University of Washington\\
    Seattle, WA 98195-2700\\
  ziyuanpu@uw.edu\\
  \hfill\break%
  \textbf{Yinhai Wang, Ph.D.}\\
  Professor\\
  Department of Civil and Environmental Engineering\\
 University of Washington\\
    Seattle, WA 98195-2700\\
  yinhai@uw.edu
}

\WordsPerTable{250}

\TotalWords{4165}

\begin{document}
\maketitle

\section{Abstract}

This study proposes to find the most appropriate transport modes with awareness of user preferences (e.g., costs, times) and trip characteristics (e.g., purpose, distance). The work was based on real-life trips obtained from a map application. Several methods including gradient boosting tree, learning to rank, multinomial logit model, automated machine learning, random forest, and shallow neural network have been tried. For some methods, feature selection and over-sampling techniques were also tried. The results show that the best performing method is a gradient boosting tree model with synthetic minority over-sampling technique (SMOTE). Also, results of the multinomial logit model show that (1) an increase in travel cost would decrease the utility of all the transportation modes; (2) people are less sensitive to the travel distance for the metro mode or a multi-modal option that containing metro, i.e., compared to other modes, people would be more willing to tolerate long-distance metro trips. This indicates that metro lines might be a good candidate for large cities.

\hfill\break%
\noindent\textit{Keywords}: Transportation Mode Choice, Recommendation System, Map Navigation, Lightgbm
\newpage

\section{Introduction}
Transport modes, such as walking, cycling, automobile, public transit, are means for traveling from an origin to a destination \cite{wang2010transportation}. Transportation mode recommendation refers to the effort of finding the most appropriate transport tools. Different from the existing and popular transportation recommendation methods on navigation applications which only provide routes in one transportation mode regardless of users' preferences, this study aims to investigate the context-aware multi-modal transportation recommendation problem.

\begin{figure}[!h]
    \centering
    \includegraphics[width=0.9\textwidth]{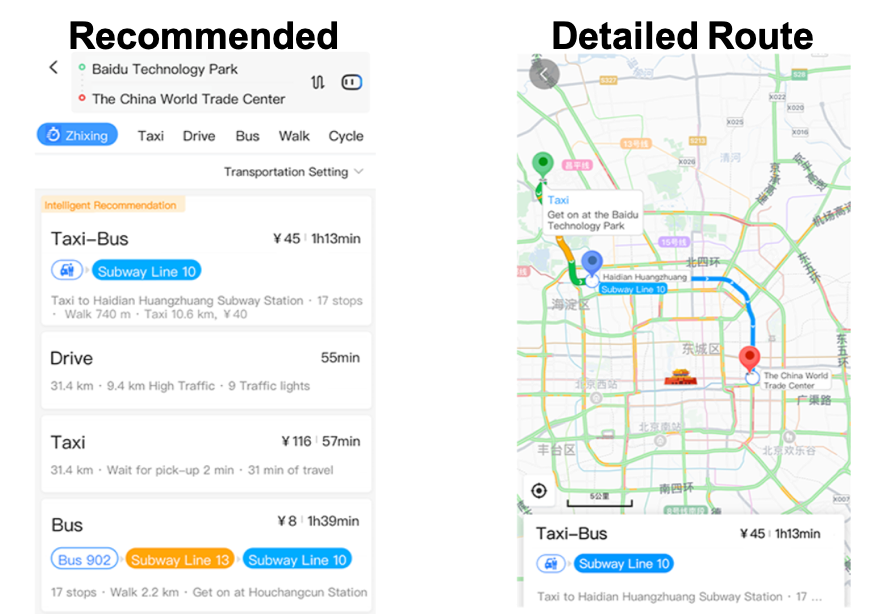}
    \caption{An example of user interfaces of context-aware multi-modal transportation recommendation service on Baidu Maps. The first recommended plan is 26.3\% faster than the pure bus plan and 61.2\% cheaper than the pure taxi plan \cite{Baidu}}
    \label{fig:disp}
\end{figure}

By \textbf{context-aware} \cite{adomavicius2011context}, we aim to address the fact that the transport mode preferences change over various users and spatiotemporal contexts \cite{wang2001spatio}. For example, metros are more cost-effective than taxis for most urban commuters; economically disadvantaged people may prefer cycling and walking to others for local travel if the transport options are inadequate \cite{Baidu}. By \textbf{multi-modal}, we intend to address the limitation that existing transportation recommendation solutions only consider routes in one transportation mode. Imagine a scenario that the distance of the OD pair is relatively large, and the trip purpose is in no emergency. In this case, a cost-effective transportation recommendation that including multiple transport modes, e.g., taxi-bus, maybe more attractive, as shown in Figure \ref{fig:disp}.

In sum, context-aware multi-modal transportation refers to recommending a travel plan consists of various modes, such as walking, cycling, driving, public transit, and ride-sharing under various contexts. It can not only help users balance travel time and travel cost, but also contribute to reducing congestion, balancing traffic flow, and promoting the development of intelligent transportation systems.

Specifically, given a user $u$, an origin-destination (OD) pair $od$, and the situational context, we want to recommend the most proper transport mode $m \in  M$ for user $u$ to travel between the OD pair $od$, considering user's preferences (e.g., costs, time) reveled in their historical trip data and trip characteristics (e.g., purpose, distance) \cite{liu2019joint}. We will investigate this problem with large-scale navigation App data.

\section{Related Work}

There are a few studies related to transportation mode recommendations. \citep{liu2019joint} proposed to recommend the most appropriate transport mode $m \in M$ for the user $u$ to travel between the OD pair $od$. Specifically, they first extracted a multi-modal transportation graph from large-scale map query data to describe the concurrency of users, OD pairs, and transport modes. Then, they developed embeddings for users, OD pairs, and transport modes based on networking embedding method. Finally, they exploited the learned representations for online multi-modal transportation recommendations.

\citep{cui2018personalized} proposed to plan an optimal travel route between two geographical locations, based on the road networks and users' travel preferences. They defined users' travel behaviors from their historical Global Positioning System (GPS) trajectories and proposed two personalized travel route recommendation methods: collaborative travel route recommendation (CTRR) and an extended version of CTRR (CTRR+). The main drawback of this study is that they only considered single-mode transportation settings. This is one of the main challenges we are going to address in the current study.

\citep{zhang2017taxi} proposed a method to predict destinations of a user based on their starting location and time. The proposed method employs the Bayesian framework to model the distribution of a user's destination based on his/her travel histories. The main advantage is that they proposed a Bayesian framework to infer users' preferences based on historical data. However, they assume that the departure time and origin locations follow Gaussian distributions, and time and location are independent. This is might not be consistent with reality and thus is a drawback of the paper.
 
 \section{Data Description}
Data in this study came from \href{https://dianshi.baidu.com/competition/29/rule}{KDD Cup 2019} \cite{Baidu}, which was provided by Baidu Map. The training set is the Baidu Map usage data from Oct. 1st to Nov. 30th, 2018. The testing set is data from Dec. 1st to Dec. 7th. Both the training and testing data were collected in Beijing, China.

The data consists of two main parts, user features data and a set of historical user behavior data. The user behavior data includes query records, display records and click records, as shown in Figure \ref{fig:data}. A total of 500,000 query records were provided in the training data. Here is the detailed information in each data table: 

\begin{figure}[!h]
    \centering
    \includegraphics[width=1\textwidth]{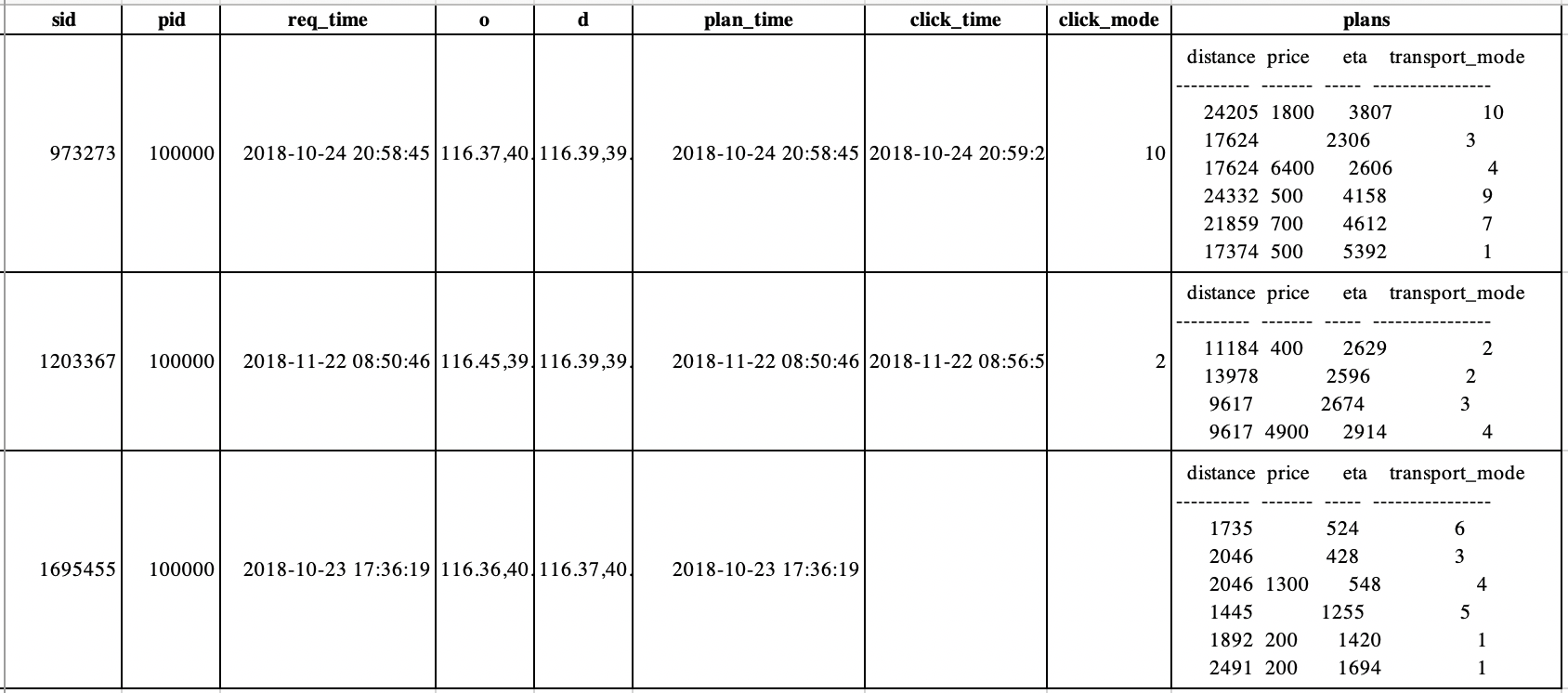}
    \caption{A preview of the user behavior data.}
    \label{fig:data}
\end{figure}

\textbf{Query Records. } A query record represents one route search from a user. Each query record consists of a session ID, a profile ID, a time stamp, and the coordinates of the origin-destination points.

\textbf{Display Records. } A display record is the list of routes generated by Baidu Maps shown to the user. Each display record consists of a session ID, a timestamp and a list of route plans. Each display plan consists of the transport mode (1 to 11), the estimated route distance (m), the estimated time of arrival (ETA) (s), the estimated price (RMB cent) and the display rank in the display list. There are 11 transport modes in total. A transport mode could be unimodal (e.g., drive, bus, cycle) or multi-modal (e.g., taxi-bus, cycle-bus). For the privacy issue, the exact meanings of the 11 transportation modes are not provided. 

\textbf{Click Records. } A click record is the specific transportation plan clicked by the user out of the whole list. A record contains a session ID, a time stamp, and the first clicked transport mode in the display list by the user.

\textbf{User Features. } User profile features reflect individual preference on transport modes. The user of each session is associated with a set of user attributes via a profile ID. Each profile record consists of a profile ID, a set of one hot encoded user profile dimensions. For the privacy issue, the real-world meaning of each attribute is not given. Also, users with the same attributes are merged, sharing the same user profile ID. For example, with gender and age attribute considered, two males of age 35 are identified as the same user in the dataset.

\section{Exploratory Data Analysis}
This section provides some initial findings and summary statistics for the dataset. In this study, the exact meaning of the 1 to 11 transportation modes is not provided, and the only features for each transportation mode are distance, time, and price. For transportation mode recommendation, these provided features are not enough because many factors that are closely related to people's mode choice behavior are ignored, such as riding comfort and time reliability. Therefore, we tried to infer the meaning of the provided transportation modes based on traveling price and speed derived from the raw data.

\begin{figure}[!h]
\begin{subfigure}{\linewidth}
\centering
\includegraphics[width=0.9\textwidth]{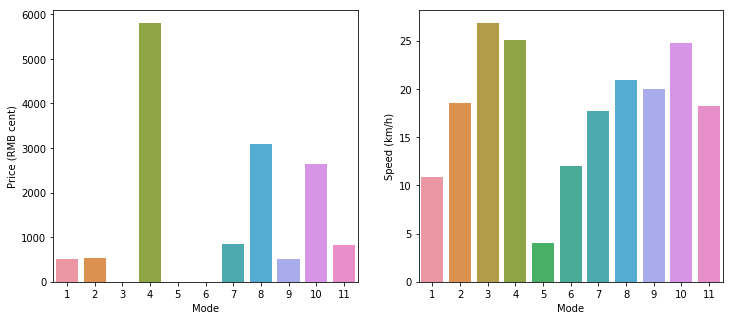}
\caption{Mean price and traveling speed for each transportation mode.}
\label{fig:sub1}
\end{subfigure}%

\medskip

\begin{subfigure}{\linewidth}
\centering
\includegraphics[width=0.9\textwidth]{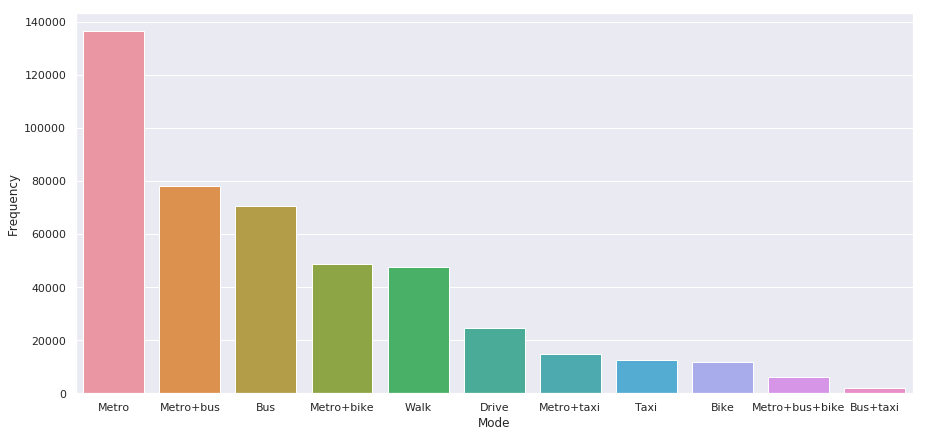}
\caption{Frequency of mode choice in ground truth data.}
\label{fig:sub2}
\end{subfigure}
\caption{Data characteristics.}
\label{fig:test}
\end{figure}

Figure \ref{fig:sub1} presents the mean price and mean speed information for each transportation mode. 
Based on common sense and Baidu Map Apps, we can make the following inferences:
\begin{itemize}
    \item Mode 3 is driving by one's own vehicle because the price is zero, and the speed is the highest. Mode 4 is by taxi because of its high price and similar speed with driving.
    \item Mode 5 and 6 are walking and biking respectively because they are free and have low speeds.
    \item Mode 1 and 2 should be transit bus and metro respectively because they are cheap, popular and with lower speeds compared to driving. 
    \item Similarly, we can identify that mode 7 is metro-bus; mode 8 is bus-taxi, mode 9 is metro-bike, mode 10 is metro-taxi, and mode 11 is metro-bus-bike.
\end{itemize}

These inferences will be valuable for the upcoming analysis because they provide additional information about our prediction targets. 


Figure \ref{fig:sub2} shows the frequency of people's mode choices in the ground truth data (i.e., users' click mode). We can see that Metro, Metro+bus, and Bus are the most frequent modes people clicked, indicating that public transportation like transit and metro serves the majority of travelers. 


\begin{figure}[!h]
    \centering
    \includegraphics[width=0.9\textwidth]{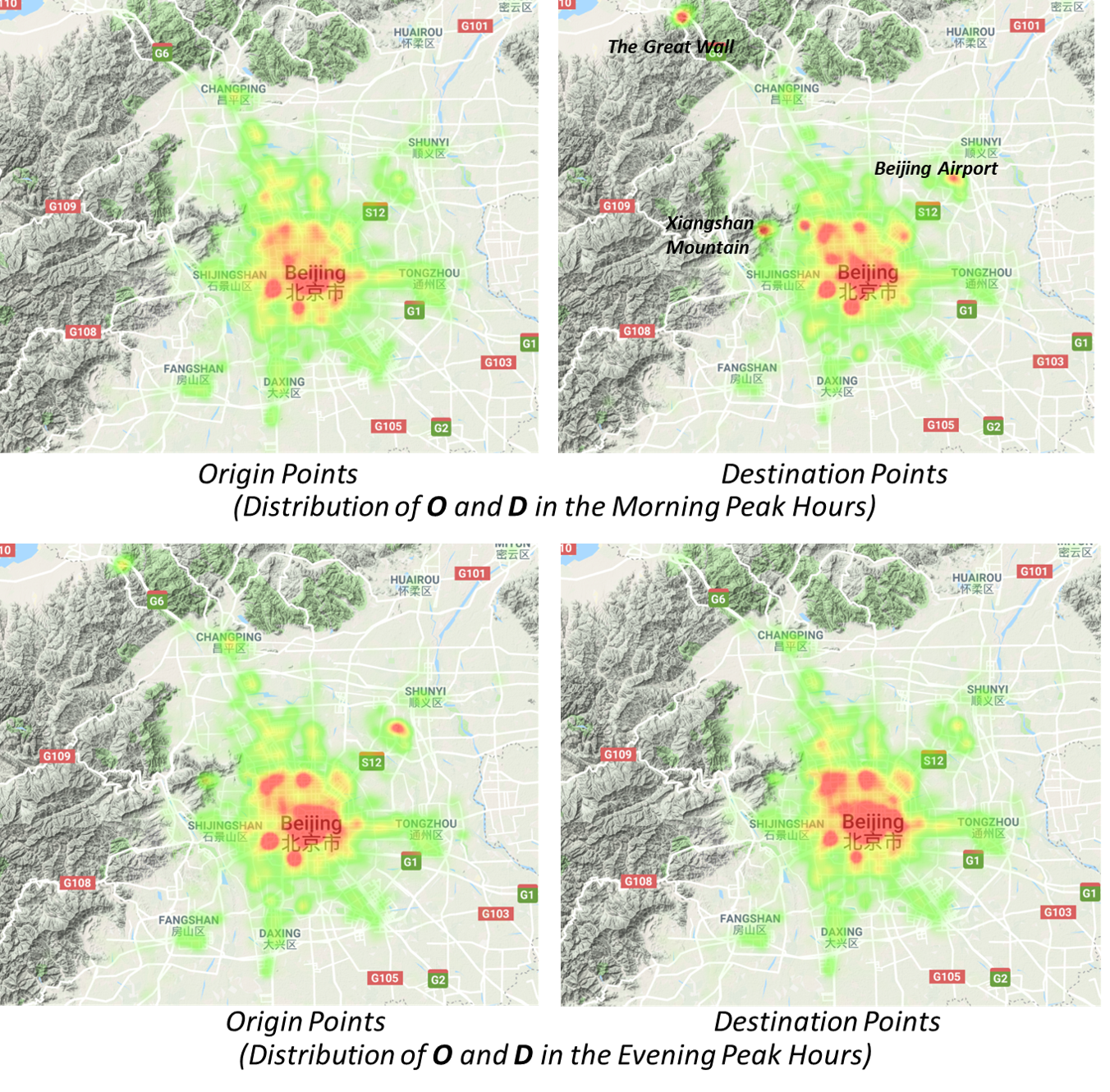}
    \caption{Heatmap of Origin and Destination points during Peak Hours.}
    \label{fig:heat}
\end{figure}

Based on the data, we can explore some activity features of the map users. By drawing heatmaps of origin (O) and destination (D) distributions during morning and evening peak hours, as shown in Figure \ref{fig:heat}, several interesting facts were found. There are some specific trip attraction points in the morning which are identified as the Beijing Airport, the Great Wall and the Xiangshan Mountain. During the evening peak hours, more users are traveling from the airport instead. Also, the suburban area has more origin points in the morning and more destination points in the evening. 

\section{Method}
We tried several methods for this problem (see Table \ref{table:results}), including gradient boosting tree methods \cite{friedman2002stochastic}, learning to rank \cite{burges2010ranknet, burges2005learning}, multinomial logit model, automated machine learning \cite{feurer2015efficient}, random forest \cite{liaw2002classification}, and shallow neural network. Feature selection and over-sampling techniques were also tried. Limited by the space, only the gradient boosting tree method and the multinomial logit model will be introduced in detail here.

\subsection{Gradient Boosting Tree}
Our main method is a gradient boosting tree model based on Lightgbm \cite{ke2017lightgbm}, treating the recommendation as a multi-class classification problem. Lightgbm is a gradient boosting framework that uses tree-based learning algorithms. It is designed to be distributed and efficient with faster training speed and higher efficiency.

Basically, we first did feature engineering to generate features for the query records, treating them as $X$. Then we took users' click mode as $y$ and did multi-class classification training and prediction. The final model included 304 features and can be summarized as follows:
\begin{itemize}
    \item \textbf{Request time features:} which hour and which weekday.
    \item \textbf{Location features:} longitudes and latitudes of the OD pairs.
    \item \textbf{Transport mode features:} distance, price, time, price*time and speed for each candidate mode; descriptive statistics of distance, price, time, speed for all the modes in each plan, like the maximum, minimum, mean, and standard deviation of distances; which mode ranked first, which mode had longest and shortest travel distances respectively; which mode took the longest and shortest time respectively, which mode had the lowest and highest price respectively.
    \item \textbf{Metro and bus station features:} distances to the 5 nearest bus/metro stations for the origin and destination locations; the number of bus/metro stations within 1500 meters' range.
    \item \textbf{Point of interest (POI) features:} the number of different nearby POIs for each origin and destination location. Used geohash to represent a location using a short alphanumeric string so that the 2 locations with the same string are considered as close to each. A total of 33 POI categories, including car service area, transportation ports (airports, railway stations), tourist area, shopping mall, etc., were used.
    \item \textbf{Location frequency features:} use geohashing to discretize the area and count the visit frequency for each origin and destination area. 
    \item \textbf{Important location features:} the distance between origin/destination location and the important trip attraction locations (Beijing Airport, the Great Wall, Xiangshan Mountain, etc.) identified through visit heat maps described in Section 3.
    \item \textbf{Profile features:} binary user profile features. 
\end{itemize}

\begin{figure}[!h]
    \centering
    \includegraphics[width=0.9\textwidth]{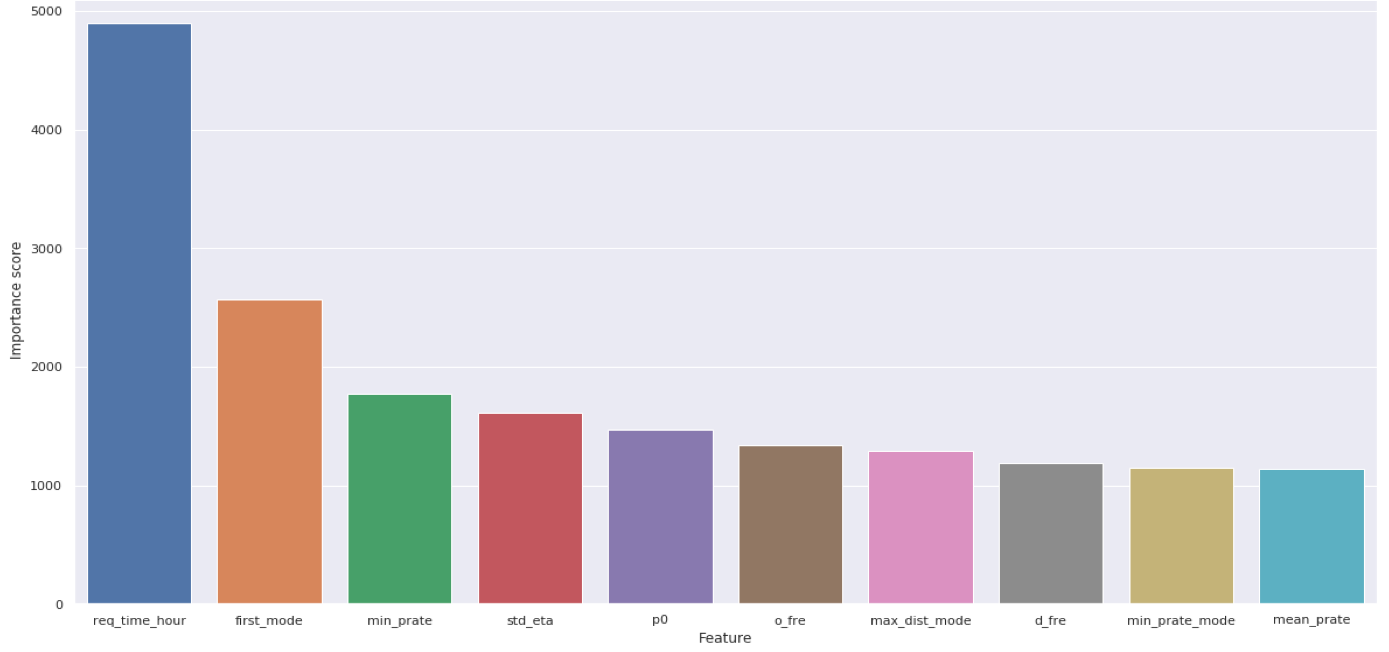}
    \caption{Top 10 important features.}
    \label{fig:fea}
\end{figure}

The top 10 important features in the final Lightgbm model are shown in Table \ref{fig:fea}, with their importance scores.

\begin{figure}[!h]
    \centering
    \includegraphics[width=1\textwidth]{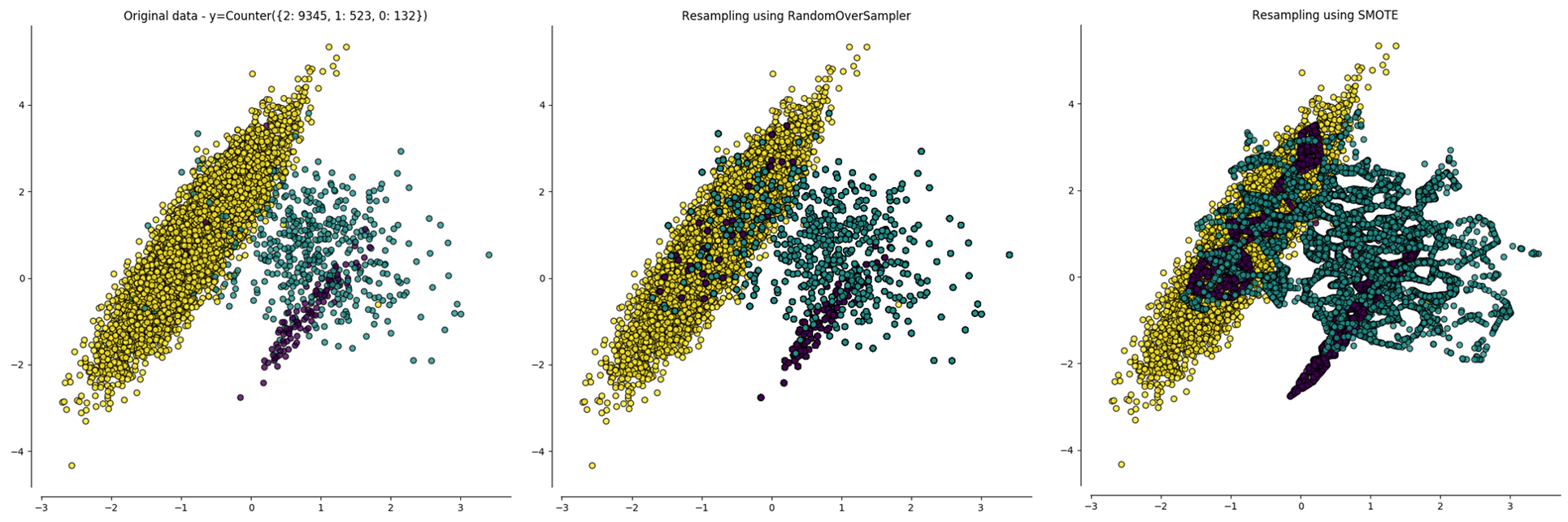}
    \caption{Illustration of SMOTE, the leftmost is the original data, the middle is results of random over-sampling, and the rightmost is the results of SMOTE.}
    \label{fig:smote}
\end{figure}

One important characteristic of this problem is the unbalanced data distribution of the clicked transport modes. With an initial analysis, we found that the model's performance on each mode is strongly correlated with the sample size of each mode: the frequently clicked modes have higher accuracy than the less frequent ones. To address this issue, we utilized an over-sampling technique called Synthetic Minority Oversampling Technique (SMOTE) \citep{chawla2002smote}. Basically, the method uses K-nearest neighbors to generate new samples for the minority classes. As shown in Figure \ref{fig:smote}, compared to directly copy existing data samples, denoted as random over-sampling, which degrades the model's performance, the SMOTE method can improve the model's performance a lot, as shown in Table \ref{table:results}. 

The detailed hyperparameters with respect to the graident boosting tree methods are listed as follows : num of leaves = 40, max depth=8, learning rate = 0.1, subsample rate = 0.8, feature selection ratio = 0.8, min child samples = 60.

\subsection{Multinomial Logit Model (MNL)}
Multinomial logit model \cite{adams1997multidimensional} is a classification method that generalizes logistic regression \cite{hosmer2013applied} to multiclass problems, i.e. with more than two possible discrete outcomes. It is a model that is used to predict the probabilities of the different possible outcomes of a categorically distributed dependent variable if given a set of independent variables, often used in modeling travelers' mode choice behavior while doing transportation planning. In this study, several different variable sets were implemented and analyzed, as shown in Table \ref{Variables}.

\begin{table}[!h]
  \caption{MNL Model Performance of Different Variable Sets}
  \label{Variables}
  \centering
  \begin{tabular}{ll}
    \toprule
    Variables     & Log-Likelihood   \\
    \midrule
    Null &-6387.030 \\
    Intercepts only	&-5365.976\\
	Intercepts and travel time for each mode	&-5608.941\\
	Intercepts and travel distance for each mode	&-5126.127\\
	Intercepts and travel cost for all the modes	&-5293.209\\
	Intercepts, travel time for each mode and travel cost for all the modes	&-5914.367\\
	Intercepts, travel distance for each mode,cost for all the modes	&-5113.514\\
	\textbf{Travel distance for each mode, cost for all the modes}	&\textbf{-5113.514}\\

    \bottomrule
  
  \end{tabular}
\end{table}

The variable set of the final MNL model with the highest Log-Likelihood is described as below. The final model results are shown in Table \ref{tab:mnl}.

\begin{itemize}
\item Travel distance: the estimated distance between the requested origin and destination pair for each mode. The coefficient of this variable varies for different modes.\par 
\item Travel cost: the estimated price of each travel mode. The coefficient of this variable is the same across all the modes. Note that some of the modes are regarded as zero-cost in the app, including walking, cycling, and driving.  \par
\end{itemize}


\begin{table}[!h]
\centering
\caption{Final MNL Model Results}
\label{tab:mnl}
\begin{tabular}{@{}lllll@{}}
\toprule
Variable & Coefficient & Std. Err. & z & p\textgreater{}|z| \\ \midrule
Distance\_bus & -3.99e-05 & 4.57e-06 & -8.734 & \textbf{0.000} \\
Distance\_metro & 9.946e-06 & 4.79e-06 & 2.078 & \textbf{0.038} \\
Distance\_drive & -0.0001 & 5.14e-06 & -24.987 & \textbf{0.000} \\
Distance\_taxi & -0.0001 & 1.23e-05 & -11.375 & \textbf{0.000} \\
Distance\_walk & -3.195e-06 & 1.87e-05 & -0.171 & 0.864 \\
Distance\_bike & -1.311e-05 & 8.91e-06 & -1.471 & 0.141 \\
Distance\_metro+bus & -3.024e-05 & 4.37e-06 & -6.928 & \textbf{0.000} \\
Distance\_metro+bike & -6.322e-06 & 4.69e-06 & -1.348 & 0.178 \\
Distance\_metro+taxi & -2.471e-05 & 5.72e-06 & -4.320 & \textbf{0.000} \\
Distance\_metro+bus+bike & -4.526e-06 & 5.86e-06 & -0.772 & 0.440 \\
Cost\_all\_mode & -4.894e-05 & 3.71e-05 & -1.319 & 0.187 \\ \bottomrule
\end{tabular}
\end{table}

Here is a summary of the main findings for the MNL model results:
\begin{itemize}
    \item An increase in travel cost would decrease the utility of all the transportation modes.  \par
    \item The baseline mode in this model is Bus+Taxi. According to the sign of the coefficients, people are more sensitive to the travel distances of all other modes except for metro.\par 
    \item People are most sensitive to travel distance while driving and taking taxis.\par 
    \item If the multi-modal choice includes metro as a sub-mode, people will be less sensitive to the travel distance. In another way, people would be more likely to use metro service compared to other modes.
\end{itemize}

\section{Prediction Results}
In this study, weighted F1 was used as the evaluation score. The F1 score for each class is defined as

\begin{linenomath}
  \begin{flalign}
    &F_{1, \text {Class}_{i}}=\frac{2 \text {precision} * \text {recall}}{\text {precision}+\text {recall}} \\
    &Precision =\frac{\text { truepositives }}{\text {truepositives}+\text {falsepositives}}\\
    &Recall=\frac{\text {truepositives}}{\text {truepositives}+\text {falsenegatives}}
  \end{flalign}
\end{linenomath}\\

As shown in the following equations, the final weighted F1 is calculated by considering the weight of each class. The weight is calculated by the ratio of true instances for each class.

\begin{linenomath}
  \begin{equation} \label{eq:2}
 F_{1, \text {weighted}}=w_{1} F_{1, \text { class }_{1}}+w_{2} F_{1, \text { Class }_{2}}+\cdots+w_{k} F_{1, \text { Class }_{k}}
\end{equation}
\end{linenomath}\\

We used data from Oct. 1st to Nov. 23th, 2018 as training data, and data from Nov. 24th to Nov. 30, 2018 as validation data. We then used the model that performed best on the validation data to predict mode choices on test data (from Dec. 1st to Dec. 7th, 2018). The performances of different methods are shown in Table \ref{table:results}. The lightgbm method with smote over-sampling performed the best. 

\begin{table}[!h]
  \caption{Performance of Different Models on Testing Data}
  \label{table:results}
  \centering
  \begin{tabular}{lllll}
    \toprule
    Model	    &Weighted F1 score    \\
    \midrule
   Lightgbm with smote oversample	& \textbf{0.6951}\\
Lightgbm	&0.6920\\
 Lightgbm with backward feature selection	&0.6912\\
 Lightgbm with pid	&0.6905\\
 Lightgbm with random oversample	&0.6884\\
 Lightgbm learning to rank	&0.6883\\
Xgboost \cite{chen2016xgboost}	&0.6877\\
Random forest \cite{breiman2001random}	&0.6851\\
Auto ML \cite{Guyon:AutoML:2015}	&0.6836\\
Catboost \cite{DBLP:journals/corr/DorogushGGKPV17}	&0.6835\\
Xgboost learning to rank	&0.6661\\
Train with single pid 
(if not have enough data, grouped with others)	&0.6645\\
Multinomial logit model	&0.4971\\
Shallow neural network (6 layers)	&0.0178\\

    \bottomrule
     
  \end{tabular}
\end{table}
 
The performance of our main method (Lightgbm with smote over-sampling) on different transportation modes is shown in Table  \ref{table:performance}. The metro mode has the highest performance maybe because it has more data samples. We can see that prediction on driving, taxi, and biking should be improved in the future because of their extremely low F1 scores. With this main method, our online testing F1 score is 0.6951, ranking top 7\% on the final leaderboard.

\begin{table}[!h]
  \caption{Lightgbm Model Performance for Each Transportation Mode on Validation Data}
  \label{table:performance}
  \centering
  \begin{tabular}{lllll}
    \toprule
    Mode     &Sample ratio    &F1 score     &Precision     &Recall    \\
    \midrule
   No click	&0.0874	&0.3552	&\textbf{0.9453} &	0.2187\\
Bus	&0.1446	&0.6804	&0.6372	&0.7299\\
Metro	&\textbf{0.3133}	&\textbf{0.9019}	&0.8543	&\textbf{0.9551}\\
Drive	&0.0446	&0.1424	&0.3997	&0.0867\\
Taxi 	&0.0245	&0.0829	&0.1836	&0.0535\\
Walk	&0.0976	&0.8452	&0.7859	&0.9143\\
Bike	&0.0199	&0.2187	&0.2529	&0.1927\\
Metro+bus &0.1779	&0.7883	&0.7112	&0.8840\\
Bus+taxi  	&0.0046	&0.3288	&0.2568	&0.4567\\
Metro+bike	&0.0499	&0.5148	&0.5803	&0.4626\\
Metro+taxi	&0.0285	&0.5424	&0.4643	&0.6521\\
Metro+bus+bike	&0.0072	&0.4187	&0.3731	&0.4771 \\

    \bottomrule
     \multicolumn{5}{c}{ Validation F1 score for all modes: \textbf{0.693168348}
} 
  \end{tabular}
\end{table}



\section{Summary and Discussion}
This study aims to find the most appropriate transport modes with awareness of user preferences (e.g., costs, times) and trip characteristics (e.g., purpose, distance). Several methods including gradient boosting tree, learning to rank, multinomial logit model, automated machine learning, random forest, and shallow neural network have been tried, and the best performing method is a gradient boosting tree (Lightgbm) model with SMOTE over-sampling. 

The Lightgbm method and the multinomial logit (MNL) model were presented in detail. In terms of accuracy, the lightgbm model is much better than the logit one. However, the logit model can provide us more about people's mode choice behavior. Some implications derived from the logit model are: (1) an increase in travel cost would decrease the utility of all the transportation modes; (2) people are less sensitive to the increase of travel distances for the metro mode or a multi-modal option that containing metro, i.e., compared to other modes, people would be more willing to tolerate long-distance metro trips. This indicates that metro lines might be a good candidate for large cities.

Below are some interesting points that deserve further discussions.
\begin{itemize}
    \item We are not sure the upper-bound limit of the F1 score for this problem. Maybe human's behavior is too hard to predict, and the F1 score is impossible to be larger than $80\%$.
    \item Oversampling gives a stable boosting for the model performance. This indicates that data or data imputing might be a major bottleneck for model performances.
    \item Individual heterogeneity does exist.
    \item Things tried but not having significant effects (or even making the performance worse):
    \begin{itemize}
\item Conduct principal component analysis (PCA) \cite{jolliffe2011principal} before learning
\item Expand the plan list and do binary classification
\item Adding class weights for unbalanced data
\item Incorporate historical mode choice probability as features
\item Scale the data
\item Normalize the data
\item Scale the time, price, distance to 0~1 within the candidate plans
\item Support vector machine (SVM) \cite{scholkopf2001learning}: too slow
\item Train with individual's data. It turned out that the behavior of some people (or a group of different people) are just too hard to predict \cite{huang2019multi, zhu2018human, zhu2019safe}
\item Adding weather as features
  \end{itemize}
  \item More combination of the variables can be tried, like interaction terms, transformation of the variables (e.g. $log(X)$, $X^2$ instead of $X$). 
  \item Embedding learning method for graph \cite{yang2016revisiting} can also be tried in future work. This will enable a better representation of transportation modes, users, and OD pairs so that classification models can have better performances.
\end{itemize}

\section{Acknowledgements}
The authors would like to thank the Pacific Northwest Transportation Consortium (PacTrans) at Regional University Transportation Center (UTC) for Federal Region 10, for funding this research. Thanks to Baidu for providing the research data, and Tim Althoff for providing constructive suggestions. 

\section{AUTHOR CONTRIBUTIONS}
The authors confirm contribution to the paper as follows: study conception and design: Hu, Zhu, Wang; data preprocess: Zhu, Hu, Yang, Pu; analysis and interpretation of results: Hu, Zhu, Yang, Pu; draft manuscript preparation: Hu, Zhu, Yang, Pu. All authors reviewed the results and approved the final version of the manuscript.

\newpage

\bibliographystyle{trb}
\bibliography{trb_template}
\end{document}